\documentclass[prd,aps,showpacs,nofootinbib,eqsecnum,onecolumn,groupedaddress,amssymb]{revtex4}
\usepackage{amsmath}
\usepackage{amssymb}
\usepackage{amsfonts}
\usepackage{graphicx,bm}
\usepackage{dcolumn}
\usepackage{color,amsxtra}
\usepackage{epsf}
\usepackage{enumerate}
\usepackage{hhline}
\usepackage{array}
\usepackage{tabularx}
\usepackage{subfigure}
\usepackage{fancyhdr}
\usepackage{mathrsfs}

\newcommand{\be}{\begin{equation}}
\newcommand{\ee}{\end{equation}}
\newcommand{\bea}{\begin{eqnarray}}
\newcommand{\eea}{\end{eqnarray}}
\newcommand{\beaa}{\begin{eqnarray*}}
\newcommand{\eeaa}{\end{eqnarray*}}

\def\be{\begin{equation}}
\def\ee{\end{equation}}
\def\bea{\begin{eqnarray}}
\def\eea{\end{eqnarray}}

\begin{document}

\title{Inflationary universe from higher-derivative quantum gravity}
\author{
R.~Myrzakulov$^{1}$,
S.~D.~Odintsov$^{2, 3, 4}$
and
L.~Sebastiani$^{1}$
}
\affiliation{
$^1$Department of General \& Theoretical Physics and Eurasian Center for
Theoretical Physics, Eurasian National University, Astana 010008, Kazakhstan\\
$^2$Consejo Superior de Investigaciones Cient\'{\i}ficas, ICE/CSIC-IEEC,
Campus UAB, Facultat de Ci\`{e}ncies, Torre C5-Parell-2a pl, E-08193
Bellaterra (Barcelona), Spain\\
$^3$Instituci\'{o} Catalana de Recerca i Estudis Avan\c{c}ats
(ICREA), Barcelona, Spain\\
$^4$ Inst.of Physics, Kazan Federal Univer.,  420008 Kazan and  Tomsk 
State Pedagogical Univer., 634061 Tomsk, Russia\\ }


\begin{abstract}
We consider higher-derivative quantum gravity where
   renormalization group improved effective action beyond one-loop
approximation is derived. Using this effective action, the
quantum-corrected FRW equations are analyzed. De Sitter universe solution
is found. It is demonstrated that such de Sitter inflationary universe is
instable. The slow-roll inflationary parameters are calculated. The
contribution of renormalization group improved Gauss-Bonnet term to
quantum-corrected FRW equations as well as to instability of de Sitter
universe is estimated. It is demonstrated that in this case the spectral
index and tensor-to-scalar ratio are consistent with Planck data.
\end{abstract}

\hspace{13.1cm}

\maketitle

\def\thesection{\Roman{section}}
\def\theequation{\Roman{section}.\arabic{equation}}

\section{Introduction}
Recent more precise observational  WMAP data \cite{WMAP} as well as corrected
Planck constraints \cite{Planck} increased the interest to the theoretical
models for inflationary universe. There is large variety of the inflationary
models (for review, see, for instance, Refs.~\cite{book}) which may comply with
observational data, at least, up to some extent (see also Ref.~\cite{Bi} about BICEP
experiment).

In fact, during last years there was much activity in the account of quantum
effects of General Relativity in the construction of inflationary universe (for
the introduction and review, see Ref.~\cite{Rich}). Furthermore, recent study
\cite{Ser} indicates that quantum effects of specific models of
(non-renormalizable) higher-derivative $F(R)$-gravity may give consistent
inflation which complies with Planck data. The next natural step is extension
of quantum-corrected inflationary scenario for multiplicatively-renormalizable
higher derivative gravity (for a general review, see Ref.~\cite{B}). The very
interesting attempt in this direction has been recently made in Ref.~\cite{Sa}.
Note that being multiplicatively-renormalizable one, higher-derivative 
quantum gravity is based on the use of higher-derivative propagator. As a 
result, such theory eventually leads to problem with unitarity what is 
related with well-known Ostrogradski instability of higher-derivative 
theories.
In fact, there were made some attempts to resolve this problem with the 
proposal that unitarity maybe restored at the non-perturbative level.
However, there is no complete proof of non-perturbative restoration of 
unitarity. Hence, so far this theory maybe considered as effective theory 
teaching us different general aspects of quantum gravity.

The purpose of the current work is the study of the inflationary universe in
general higher-derivative quantum gravity~\cite{B}. Making use the fact that
one-loop beta-functions of such theory are well-known and their asymptotically
free regime is well investigated, we apply renormalization group (RG)
considerations to get RG improved effective action in general higher-derivative
gravity. This technique is well-developed in quantum field theory in curved
spacetime~\cite{El}. It permits to get the effective action beyond one-loop
approximation, making sum of all leading logs of the theory.

The paper is organized as follows. In Section {\bf 2}, we present the
renormalization-group improved effective action of
multiplicatively-renormalizable higher-derivative gravity. In order to do so,
the one-loop effective  coupling constants are used. Subsequently, the
quantum-corrected equations of motion are derived on the flat
Friedmann-Robertson-Walker space-time. In Section {\bf 3}, using the asymptotic
behaviour of the gravitational running constants, de Sitter  inflationary
universe is constructed. The asymptotically-free regime is discussed in detail.
Section {\bf 4} is devoted to the study of the dynamics of such
quantum-corrected inflation. It is shown that  de Sitter space is unstable and
can lead to a large amount of inflation. Slow-roll conditions are   discussed
and the expressions for slow-roll parameters are found. In Section {\bf 5}, we
consider the contribution from total derivative and surface terms (topological
Gauss-Bonnet
term and dalambertian of the curvature) to RG improved effective action. It is
demonstrated that with these terms the spectral index can be
compatible with Planck data. Conclusions and final remarks are given in Section
{\bf 6}.

\section{Renormalization-group improved effective action and quantum-corrected
FRW equations}

In this section we start from the general action of higher-derivative gravity
which is known to be multiplicatively-renormalizable theory (see Ref.~\cite{B}
for general introduction and review).
The starting action has the following form\footnote{
Note that higher-derivative theory of the type of (\ref{azione0}) as well as other
higher-derivative modified gravities may   even pass solar system tests,
for instance, due to chameleon scenario
~\cite{test} and so on.
}:
\begin{equation}
I=\int_\mathcal{M} d^4 x\sqrt{-g}\left(\frac{R}{\kappa^2}-\Lambda+a
R_{\mu\nu}R^{\mu\nu}+b R^2+c R_{\mu\nu\xi\sigma}R^{\mu\nu\xi\sigma}+d\Box
R\right)\,,\label{azione0}
\end{equation}
where $g$ is the determinant of the metric tensor $g_{\mu\nu}$, $\mathcal{M}$
is the space-time manifold, $R\,, R_{\mu\nu}\,,R_{\mu\nu\xi\sigma}$ are the
Ricci scalar, the Ricci tensor and the Riemann tensor, respectively, and
$\Box\equiv g^{\mu\nu}\nabla_{\mu}\nabla_{\nu}$ is the
covariant d'Alembertian, ${\nabla}_{\mu}$ being the covariant derivative
operator associated with the metric $g_{\mu \nu}$.
Here,
$\kappa^2>0$, $\Lambda\,,a\,,b\,,c$ and $d$ are constants which characterize
the gravitational interaction. The above lagrangian contains some terms not
important in four dimensions. First of all, we note that $\Box R$ is a surface
term which does not give any contribution to the dynamical equations. Second,
we have
\begin{equation}
R_{\mu\nu}R^{\mu\nu}=\frac{C^2}{2}-\frac{G}{2}+\frac{R^2}{3}\,,\quad
R_{\mu\nu\xi\sigma}R^{\mu\nu\xi\sigma}=2C^2-G+\frac{R^2}{3}\,,
\end{equation}
where $G$ and $C^2$ are the Gauss-Bonnet term and the ``square'' of the Weyl
tensor,
\begin{equation}
G=R^2-4R_{\mu\nu}R^{\mu\nu}+R_{\mu\nu\xi\sigma}R^{\mu\nu\sigma\xi}\,,\quad
C^2=\frac{1}{3}R^2-2R_{\mu\nu}R^{\mu\nu}+R_{\xi\sigma\mu\nu}R^{\xi\sigma\mu\nu}\,.\label{Gauss}
\end{equation}
The Gauss-Bonnet term is a topological invariant in four dimensions, and we can
drop
    it from the action. Thus, we can rewrite the higher derivative terms with the
help  of the Weyl squared tensor.

Let us express the constants which appear in the starting action in terms of
more convenient coupling constants which stress that the theory under
consideration is asymptotically-free one. In order to do it, we follow the
notations of Ref.~\cite{B}.
To take into account quantum gravity effects we use the
renormalization-group (RG) improved effective action. The calculation of RG
improved effective action has been developed in multiplicatively-renormalizable
quantum field theory in curved spacetime. In general terms, this technique is
described in detail in Refs.~\cite{B,El}. Recently, RG improved scalar
potential
in curved spacetime has been applied in the study of inflation \cite{rginfl}.
In the simplest version \cite{El}, RG improved effective action follows from
the solution of RG equation applied to complete effective action of the
multiplicatively renormalizable theory. The final result is very simple: one
has to replace constants in the classical action by one-loop effective coupling
constants where corresponding RG parameter is defined as log term of
chacteristic mass scale in the theory.

Applying the above considerations to higher-derivative quantum gravity,
one can get RG improved effective action as the following:
\begin{equation}
I=\int_\mathcal{M}d^4\sqrt{-g}\left[\frac{R}{\kappa^2(t')}-\frac{\omega(t')}{3\lambda(t')}R^2+\frac{1}{\lambda(t')}C^2-\Lambda(t')\right]\,.
\label{action}
\end{equation}
The effective coupling constants $\lambda\equiv\lambda(t')$,
$\omega\equiv\omega(t')$, $\kappa^2\equiv\kappa^2(t')$ and
$\Lambda\equiv\Lambda(t')$ obey to the one-loop RG
equations~\cite{Fradkin}
\begin{eqnarray}
\frac{d\lambda}{d
t'}&=&-\beta_2\lambda^2\equiv-\left(\frac{133}{10}\right)\lambda^2\,,\label{lambdaeq}\\
\frac{d\omega}{d
t'}&=&-\lambda(\omega\beta_2+\beta_3)\equiv-\lambda\left(\frac{10}{3}\omega^2+\frac{183}{10}\omega+\frac{5}{12}\right)\,,\label{omegaeq}\\
\frac{d\kappa^2}{d
t'}&=&\kappa^2\gamma\label{kappaeq}\equiv\kappa^2\lambda\left(\frac{10}{3}\omega-\frac{13}{6}-\frac{1}{4\omega}\right)\,,\label{kappaeq}\\
\frac{d\Lambda}{d
t'}&=&\frac{\beta_4}{\left(\kappa^2\right)^2}-2\gamma\Lambda(t')\equiv
\frac{\lambda^2}{(\kappa^2)^2}\left(\frac{5}{2}+\frac{1}{8\omega^2}\right)+\lambda\Lambda
\left(\frac{28}{3}+\frac{1}{3\omega}\right)
\label{Capitallambdaeq}\,.
\end{eqnarray}
Note that $\kappa^2(t')$ is positive defined, and in general $\lambda(t')$ and
$\Lambda(t')$ are also positive defined to have a positive contribution to the
Weyl tensor and a positive effective cosmological constant in the action; on
the other hand, $\omega(t')$ is expected to be negative to have a positive
$R^2$-term.
In the above expressions, $\beta_{2,3,4}$ and $\gamma$ correspond to~\cite{B}
\begin{equation}
\beta_2=\frac{133}{10}\,,\quad\beta_3=\frac{10}{3}\omega^2+5\omega+\frac{5}{12}\,,\quad
\beta_4=\frac{\lambda^2}{2}\left(5+\frac{1}{4\omega^2}\right)+\frac{\lambda}{3}\left(\kappa^2\right)^2\Lambda\left(20\omega+15-\frac{1}{2\omega}\right)\,,
\quad
\gamma=\lambda\left(\frac{10}{3}\omega-\frac{13}{6}-\frac{1}{4\omega}\right)\,.\label{betabeta}
\end{equation}
The RG parameter $t'$ is given by \begin{equation}
t'=\frac{t'_0}{2}\log \left[\frac{R}{R_0}\right]^2\,,
\label{tprime}
\end{equation}
where $t'_0>0$ is dimensionless constant introduced for the sake of
completeness and $R_0$ is the mass
scale for the Ricci scalar. We set $R_0$ as the value of the Ricci scalar in
the current nearly de Sitter universe ($R_0=4\Lambda$, $\Lambda$ being the
cosmological constant), such that $t'(R=R_0)=0$ today, while in the past
$0<t'(R_0<R)$. Note that the de Sitter solution of the current accelerated
expansion is a final attractor of Friedmann universe.

For Eq.~(\ref{lambdaeq}) we also have the explicit solution
\begin{equation}
\lambda(t')=\frac{\lambda(0)}{1+\lambda(0)\beta_2 t'}\,,\label{lambda}
\end{equation}
where $\lambda(0)$ is the integration constant corresponding to the value of
$\lambda$ at $t'=0$, namely $\lambda(t=t_0)\equiv\lambda(R=R_0)=\lambda(0)$.

One important remark is in order: when we introduce the effective running
constants in (\ref{azione0}), we also get a contribution from the Gauss-Bonnet
and $\Box R$ in RG improved effective action, since it is not more possible to
write the Gauss-Bonnet term like a total derivative and $\Box R$ in terms of a
flux in three dimensions. This fact will be discussed in below, but for the
moment we
    work with the simplified action. \\
\\
Let us consider the flat Friedmann-Robertson-Walker (FRW) space-time, whose
general form is given by
\begin{equation}
ds^2=- N(t)^2dt^2+a(t)^2(dx^2+dy^2+dz^2)\,,\label{metric}
\end{equation}
where $a\equiv a(t)$ is the scale factor depending on the cosmological time $t$
and $N\equiv N(t)$ is an arbitrary lapse function, which describes the gauge
freedom associated with the reparametrization invariance of the action.
For the above metric, the Ricci scalar and the square of the Weyl tensor read
\begin{equation}
R=\frac{1}{N^2}\left[6\left(\frac{\dot a}{a}\right)^2+6\left(\frac{\ddot
a}{a}\right)-6\left(\frac{\dot N}{N}\right)\left(\frac{\dot
a}{a}\right)\right]\,,\label{Ricci}
\quad C^2=0\,,
\end{equation}
where the dot denotes the derivative with respect to the cosmological time $t$.
The fact that the Weyl tensor is zero on the general form of the metric
indicates that its contribution to the action and therefore to the derivation
of the field equations of the theory is null. In fact one can write on FRW
background
\begin{equation}
\delta I_{C^2}=\frac{1}{\lambda(t')}\delta\left(\sqrt{-g}C^2\right)+
\left(\sqrt{-g}C^2\right)\delta\left(\frac{1}{\lambda(t')}\right)=\frac{1}{\lambda(t')}\delta\left(\sqrt{-g}C^2\right)\,,
\end{equation}
but
\begin{equation}
\frac{1}{\lambda(t')}\delta\left(\sqrt{-g}C^2\right)=0\,,
\end{equation}
and it is well known that the square of the Weyl tensor does not enter in the
Friedmann-like equations.

To derive the equations of motion (EOMs), we will use a method based on the
Lagrangian multiplayer~\cite{L1, L2, Monica, miolag}. If we plug the expression
for the Ricci scalar (\ref{Ricci}) into the action (\ref{action}), we get
higher derivative lagrangian theory. In order to derive a standard (first
order) lagrangian theory, we introduce a Lagrangian multiplier $\xi$
as~\cite{L1,L2},
\begin{equation}
I=\int_\mathcal{M}d^4\sqrt{-g}\left[\frac{R}{\kappa^2(t')}-\frac{\omega(t')}{3\lambda(t')}R^2-\Lambda(t)-\xi\left[R-\frac{1}{N^2}\left[6\left(\frac{\dot
a}{a}\right)^2+6\left(\frac{\ddot{a}}{a}\right)-6\left(\frac{\dot
N}{N}\right)\left(\frac{\dot a}{a}\right)\right]\right]\right]\,,
\label{action2}
\end{equation}
where we have taken into account (\ref{Ricci}). By making the derivation with
respect to $R$, one  finds
\begin{equation}
\xi=-2R\frac{\omega(t')}{3\lambda(t')}+\frac{1}{\kappa^2}-\Delta(t')\frac{d
t'}{d R}\,,\label{xi}
\end{equation}
where
\begin{equation}
\Delta(t')=\left[\frac{R}{(\kappa^2(t'))^2}\frac{d\kappa^2(t')}{d
t'}+R^2\frac{d}{d
t'}\left(\frac{\omega(t')}{3\lambda(t')}\right)+\frac{d\Lambda(t')}{d
t'}\right]\,,
\end{equation}
since it is understood that the functions
$\kappa^2(t')\,,\Lambda(t')\,,\lambda(t')$ and $\omega(t')$ depend on $R$
throught $t'$ as in Eq.~(\ref{tprime}).

Therefore, by substituting (\ref{xi}) and making an integration by parts one
obtains the (standard) Lagrangian
\begin{eqnarray}
\mathcal L(a,\dot a, N, R, \dot R)&=&
-Na^3\Lambda(t')-\frac{6\dot a^2 a}{\kappa^2(t')N}+\frac{6\dot a
a^2\dot{(\kappa^2(t'))}}{N(\kappa^2(t'))^2}
+\frac{\omega(t')}{3\lambda(t')}a^3 N
\left[R^2+\frac{12R}{N^2}\frac{\dot a^2}{a^2}
+\frac{12\dot a\dot R}{a N^2}\right]\nonumber\\&&
\hspace{-2cm}+\frac{d}{dt'}\left[\frac{\omega(t')}{3\lambda(t')}\right]\left(\frac{d
t'}{d R}\dot R\right)\frac{12Ra^2\dot a}{N}
+6a^3 N\left(\frac{R}{6}+\frac{\dot a^2}{a^2 N^2}\right)\Delta(t')\frac{d t'}{d
R}
+6\dot a\left(\frac{a^2}{N}\right)
\left[\frac{d\Delta(t')}{d t'}\left(\frac{d t'}{d
R}\right)^2+\Delta(t')\frac{d^2 t'}{d R^2}\right]\dot R\,.\nonumber\\
\label{lagrangian}
\end{eqnarray}
If we derive this Lagrangian with respect to $N(t)$ and therefore we choose the
gauge $N(t)=1$, we get
\begin{eqnarray}
0&=&
-a^3\Lambda(t')+\frac{6\dot a^2 a}{\kappa^2(t')}-\frac{6\dot a
a^2\dot{(\kappa^2(t'))}}{(\kappa^2(t'))^2}
+\frac{\omega(t')}{3\lambda(t')}a^3
\left[R^2-12R\frac{\dot a^2}{a^2}
-\frac{12\dot a\dot R}{a}\right]
-12 R a^2 \dot a\frac{d}{d
t'}\left(\frac{\omega(t')}{3\lambda(t')}\right)\left(\frac{d t'}{d R}\dot
R\right)\nonumber\\&&
+6a^3\left(\frac{R}{6}-\frac{\dot a^2}{a^2}\right)\Delta(t')\frac{d t'}{d
R}-6\dot a a^2
\left[\frac{d\Delta(t')}{d t'}\left(\frac{d t'}{d
R}\right)^2+\Delta(t')\frac{d^2 t'}{d R^2}\right]\dot R\,.
\label{NN}
\end{eqnarray}
The variation with respect to $a(t)$ leads to
\begin{eqnarray}
0&=&-3a^2\Lambda(t')+\frac{6}{\kappa^2(t')}\left(\dot a^2+2\ddot a
a\right)+\frac{6}{\kappa^2(t')}
\left(\frac{2a^2\dot{(\kappa^2(t'))}^2}{(\kappa^2(t'))^2}-\frac{2\dot a a
\dot{(\kappa^2(t'))}}{\kappa^2(t')}-\frac{a^2(\ddot{\kappa^2(t'))}}{\kappa^2(t')}\right)+\nonumber\\
&&\frac{\omega(t')}{\lambda(t')}\left(R^2a^2-4R\dot a^2-8\dot R\dot a a-8R\ddot
a a-4\ddot R a^2\right)
-24\frac{d}{dt}\left(\frac{\omega(t')}{\lambda(t')}\right)\left[\dot R a^2+R a
\dot a\right]
-12\frac{d^2}{dt^2}\left(\frac{\omega(t')}{\lambda(t')}\right)R
a^2\nonumber\\&&
+\left(3a^2 R-6\dot a^2-12a\ddot a\right)\Delta(t')\frac{d t'}{d R}
-\left(12 a \dot a \dot R+6 a^2 \ddot R\right)
\left[\frac{d\Delta(t')}{d t'}\left(\frac{d t'}{d
R}\right)^2+\Delta(t')\frac{d^2 t'}{d R^2}\right]
\nonumber\\&&
-6a^2\dot R^2
\left[\frac{d^2\Delta(t')}{d t'^2}\left(\frac{d t'}{d R}\right)^3
+3\frac{d \Delta(t')}{d t'}\left(\frac{d t'}{d R}\right)\frac{d^2 t'}{d R^2}
+\Delta(t')\frac{d^3 t'}{d R^3}\right]\,,
\label{aa}
\end{eqnarray}
where we have set $N(t)=1$ again and $d/dt\equiv \dot R(d t'/dR)d/dt'$.
Finally, the variation of the Lagrangian with respect to $R$, remembering that
$t'$ is a function of $R$, returns to be the expression in (\ref{Ricci}), and
by putting $N(t)=1$ we have
\begin{equation}
R=6\left(\frac{\dot a}{a}\right)^2+6\left(\frac{\ddot a}{a}\right)\,.\label{RR}
\end{equation}
We obtained a system of three second order equations (\ref{NN})--(\ref{RR}),
where one is redundant (in the absence of matter contributions), namely it can
be derived from the other two.

Eq.(\ref{NN}) and Eq.~(\ref{RR}) can be rewritten as
\begin{eqnarray}
0&=&
-\Lambda(t')+\frac{6 H^2}{\kappa^2(t')}-\frac{6
H}{(\kappa^2(t'))^2}\frac{d\kappa^2(t')}{d t'}\left(\frac{t_0' \dot
R}{R}\right)
+\frac{\omega(t')}{3\lambda(t')}
\left[6R\dot H
-12H\dot R\right]
-12 H\frac{d}{d t'}\left(\frac{\omega(t')}{3\lambda(t')}\right)\left(\dot R
t_0'\right)\nonumber\\&&
+6\left(H^2+\dot H\right)\Delta(t')\frac{t_0'}{R}-6 H
\left[\frac{d\Delta(t')}{d
t'}\left(\frac{t_0'}{R}\right)^2-\Delta(t')\frac{t_0'}{R^2}\right]\dot R\,,
\label{NN2}
\end{eqnarray}
\begin{equation}
R=12H^2+6\dot H\label{R2}\,,
\end{equation}
where we have introduced the Hubble parameter $H=\dot a /a$ and we have used
(\ref{tprime}) to write $d t'/dR=t_0'/R$. In the following expression, we
explicit develop Eq.~(\ref{NN2}) in terms of the functions
$\lambda(t')\,,\omega(t')\,,\kappa^2(t')$ and $\Lambda(t')$ by using the set of
equations (\ref{lambdaeq})--(\ref{Capitallambdaeq}) and Eq.~(\ref{R2}) for the
Ricci scalar,
\begin{eqnarray}
0&=&\frac{12 \omega  \left(-6 H^2 \dot H-2 H \ddot H+\dot H^2\right)}{\lambda
      }-\frac{H \lambda  t_0' \left(40 \omega ^2-26 \omega -3\right) (4 H
      \dot H+\ddot H)}{2 \kappa ^2 \omega  \left(2 H^2+\dot H\right)}+\frac{6
      H^2}{\kappa ^2}\nonumber\\
&&
-\frac{t_0'}{360 \kappa ^4 \omega ^3 \left(2
      H^2+\dot H\right)^2}
\left(H (4 H \dot H+\ddot H) \left(\lambda
      t_0 '\left(120 \kappa ^4 \omega ^3 (4 \omega +3) (2 \omega  (100 \omega
      +549)+25) \left(2 H^2+\dot H\right)^2
\right.\right.\right.
\nonumber\\&&
-2 \kappa ^2 \lambda  \omega  \left(24 H^2
      (\omega  (\omega  (20 \omega  (100 \omega +409)-2121)+210)+15)+12 \dot H
(\omega
      (\omega  (20 \omega  (100 \omega +409)-2121)+210)+15)
\right.
\nonumber\\&&
\left.\left.
+\kappa ^2 \Lambda  (4 \omega
      (1616 \omega -355)-45)\right)-15 \lambda ^2 (\omega  (2 \omega  (4 \omega
(50
      \omega +97)-25)-71)-5)\right)-180 \kappa ^2 \omega ^2 \left(2 H^2+\dot
H\right)
\nonumber\\&&
\left.
      \left(20 \kappa ^2 \omega  (4 \omega  (2 \omega +3)+1) \left(2
      H^2+\dot H\right)+\lambda  \left(-40 \omega ^2+26 \omega
      +3\right)\right)\right)+15 \omega  \left(2 H^4+7 H^2 \dot H+H
      \ddot H+\dot H^2\right)
\nonumber\\&&
\left(120 \kappa ^4 \omega ^2 (4 \omega  (2 \omega
      +3)+1) \left(2 H^2+\dot H\right)^2+4 \kappa ^2 \lambda  \omega  \left(6 H^2
      \left(-40 \omega ^2+26 \omega +3\right)+\dot H (6 (13-20 \omega ) \omega
+9)
\right.\right.
\nonumber\\&&
\left.\left.\left.
-2
      \kappa ^2 \Lambda  (28 \omega +1)\right)-3 \lambda ^2 \left(20 \omega
      ^2+1\right)\right)\right)+10 H t_0 \left(8 \omega ^2+12 \omega +1\right) (4
H
      \dot H+\ddot H)-\Lambda\,.\label{sistemone}
\end{eqnarray}
Here, $\lambda\equiv\lambda(t')$, $\omega\equiv\omega(t')$,
$\kappa^2\equiv\kappa^2(t')$ and $\Lambda\equiv\Lambda(t')$.
One should  remember that $t'$ is related to $R$ as in Eq.~(\ref{tprime}), and
only $\lambda(t')$ is given by (\ref{lambda}). Note that the above approach
suggests the consistent way to account for quantum effects of higher-derivative
gravity. Note also that different approach to take into account such quantum
effects at the inflationary universe was developed in Ref.~\cite{Sa}.

On the de Sitter solution $R_\text{dS}=12 H_\text{dS}^2$, where $H_\text{dS}$
is a constant, the system is simplified as
\begin{eqnarray}
0&=&
\frac{6 H^2}{\kappa ^2}-\frac{t_0'}{48 (\kappa ^2)^2 \omega ^2} \left(480 H^4
(\kappa^2)^2 \omega ^2 (4 \omega  (2
      \omega +3)+1)+4 \kappa ^2 \lambda  \omega  \left(6 H^2 \left(-40 \omega
^2+26 \omega
      +3\right)-2 \kappa ^2 \Lambda  (28 \omega +1)\right)
\right.
\nonumber\\&&
\left.
-3 \lambda ^2 \left(20 \omega
      ^2+1\right)\right)-\Lambda\,,
\label{dSeq}
\end{eqnarray}
where the functions $\lambda\,,\omega\,,\kappa^2$ and $\Lambda$ are assumed to
be constant and $H\equiv H_\text{dS}$.

Hence, we obtained consistent system of quantum-corrected FRW equations from RG
improved effective action corresponding to higher-derivative quantum gravity.

\section{Asymptotic behaviour of the effective  coupling constants and de
Sitter solution for inflation}

In order to solve the system (\ref{sistemone}), we need to investigate the
asymptotic behaviour of the implicitly-given effective coupling constants
$\omega(t')\,,\kappa^2(t'), \Lambda(t')$, when $t'\rightarrow\infty$, namely at
the high curvature limit ($R\rightarrow\infty$) describing inflation (see
(\ref{tprime})). Eq. (\ref{omegaeq}) has two fixed points at
\begin{equation}
\omega_1\simeq-0.02\,,\quad\omega_2\simeq-5.47\,,
\end{equation}
and the  analysis of the solution around the fixed points
$\omega(t')=\omega_{1,2}+\delta\omega(t')$, with $|\delta\omega(t')|\ll 1$,
leads to
\begin{eqnarray}
\hspace{-1cm}\frac{d\omega(t')}{d
t'}&\simeq&-\lambda(t')\left(\frac{20}{3}\omega+\frac{183}{10}\right)\vert_{\omega_{1,2}}\delta\omega(t')
-\lambda(t')^2\beta_2\left(\frac{d
t'}{d\omega(t')}\right)\left(\frac{10}{3}\omega^2+\frac{183}{10}\omega+\frac{5}{12}\right)\vert_{\omega_{1,2}}\delta\omega(t')\nonumber\\
&=&-\lambda(t')\left(\frac{20}{3}\omega+\frac{158}{5}\right)\vert_{\omega_{1,2}}\delta\omega(t')\,,\label{omegapert}
\end{eqnarray}
such that,
\begin{equation}
\omega(t')=\omega_{1,2}+\frac{c_0}{(1+\lambda(0)\beta_2 t')^{q}}\,,\quad
q=\frac{1}{\beta_2}
\left(\frac{20}{3}\omega+\frac{158}{5}\right)\vert_{\omega_{1,2}}\,,\quad
|c_0|\ll 1\,,\label{inter}
\end{equation}
where $c_0$ is a constant and we have introduced $\lambda(t')$ as in
(\ref{lambda}).
We immediatly see that $q\simeq2.37$ for $\omega_1$ rendering the solution
stable when $t'\rightarrow\infty$, but for $\omega_2$ one gets $q\simeq-0.37$
and the solution is unstable when $t'\rightarrow\infty$. Thus, we expect that
for large values of $t'$ the function $\omega(t')$ tends to the attractor
$\omega_1$. Since between $\omega_1$ and $\omega_2$ the derivative
$d\omega(t')/d t'$ with $0<\lambda(t')$ is positive, $\omega(t')$ grows up with
$t'$ and approaches to $\omega_1$ being $\omega(t')<\omega_1$.
When $\omega_2<\omega(t')<\omega_1$ we may estimate from (\ref{omegapert}),
\begin{equation}
\frac{d\omega(t')}{d
t'}=-\frac{\lambda(t')}{2}\left(\frac{20}{3}\right)\left(\omega_1-\omega_2\right)\delta\omega(t')\,.
\end{equation}
Therefore, the solution (\ref{inter}) is rewritten as (see third Ref.
in~\cite{Fradkin}),
\begin{equation}
\omega(t')=\omega_{1}+\frac{c_0}{(1+\lambda(0)\beta_2 t)^{p}}\,,\quad
p=\left(\frac{10}{3}\right)\frac{(\omega_1-\omega_2)}{\beta_2}\simeq
1.36\,,\quad |c_0|\ll1\,.\label{omega1}
\end{equation}
Note that related study for the behaviour of above dimensionless coupling
constants in relation with dimensional transmutation is given in
Ref.~\cite{Ei}.

In order to study the behaviour of $\kappa^2(t')$ and $\Lambda(t')$, we
introduce
\begin{equation}
\tilde\Lambda(t')=(\kappa^2(t'))^2\Lambda(t')\,,
\end{equation}
and Eq.~(\ref{Capitallambdaeq}) with Eq.~(\ref{kappaeq}) lead to
\begin{equation}
\frac{d\tilde\Lambda(t')}{d t'}=
\beta_4\equiv\frac{\lambda(t')^2}{2}\left(5+\frac{1}{4\omega(t')^2}\right)+\lambda(t')\tilde\Lambda(t')\left(\frac{20}{3}\omega(t')+5-\frac{1}{6\omega(t')}\right)\,.
\end{equation}
In the asymptotic limit $\omega(t')\simeq\omega_1$ we get
\begin{equation}
\tilde\Lambda=-\frac{3\lambda(0)(1+20\omega_1^2)}{4\omega_1(1+\lambda(0)\beta_2
t')(-1+30\omega_1+6\beta_2\omega_1+40\omega_1^2)}
+\tilde\Lambda_0(1+\lambda(0)\beta_2 t')^{W/\beta_2}\,,\quad
W=\frac{20}{3}\omega_1+5-\frac{1}{6\omega_1}=13.2\,.
\end{equation}
As a consequence,
\begin{equation}
\tilde\Lambda(t')\simeq \tilde\Lambda_0(1+\lambda(0)\beta_2 t')^{W/\beta_2}\,,
\end{equation}
where the constant $\tilde\Lambda_0$ is assumed to be positive. On the other
side, from Eq.~(\ref{kappaeq}) we have at $\omega(t')\simeq\omega_1$,
\begin{equation}
\kappa^2(t')\simeq\kappa_0^2(1+\lambda(0)\beta_2 t')^{Z/\beta_2}\,,\quad
Z=\left(\frac{10}{3}\omega_1-\frac{13}{6}-\frac{1}{4\omega_1}\right)\simeq
10.27\,,
\end{equation}
such that finally
\begin{equation}
\Lambda(t')\simeq\frac{\tilde\Lambda_0}{(\kappa_0^2)^2}(1+\lambda(0)\beta_2
t')^{X/\beta_2}\,,\quad X=(W-2Z)\simeq-7.34\,.
\end{equation}
Let us summarize the results. From the investigation of the asymptotic region,
we can derive the effective running coupling constants of the model
(\ref{action}) as
\begin{equation}
\lambda(t')=\frac{\lambda(0)}{(1+\lambda(0)\beta_2 t')}\,,\quad
\omega\simeq\omega_{1}+\frac{c_0}{(1+\lambda(0)\beta_2 t')^{1.36}}\,,\quad
\kappa^2(t')\simeq\kappa_0^2(1+\lambda(0)\beta_2 t')^{0.77}\,,\quad
\Lambda(t')\simeq\Lambda_0\frac{1}{(1+\lambda(0)\beta_2
t')^{0.55}}\,.\label{setuno}
\end{equation}
Here, $\Lambda_0=\tilde\Lambda_0/(\kappa_0^2)^2$ and $|c_0|\ll |\omega_1|$, and
we will omit its contribution at large $t'$. One remark is in order. In
principle these expressions correspond to the behaviour of the coupling
constants in the high energy limit, when $t'\rightarrow\infty$ and $R_0\ll R$,
$R_0$ being the Ricci scalar at the present time, and they are valid as soon as
$\omega(t')$ is close to $\omega_1$.
However, we may assume that the structure of the coupling constants keeps the
same form at every epoch, since in fact out of inflation the curvature of the
universe drastically decreases, $t'\rightarrow 1$, and the coupling constants
are expected to be constant: in fact, we can consider $\omega(t')$ sufficiently
close to $-\omega_1$ at every time, namely we will not consider the additional
corrections
at small curvature.
In particulary, at the present de Sitter epoch with $R=R_0$ and $t'_0=0$ (see
Eq.~(\ref{tprime}) and the comment below) we must find
\begin{equation}
\kappa^2(t'_0)\equiv\kappa^2_0=\frac{16\pi}{M_{Pl}^2}\,,\quad
\Lambda(t'_0)\equiv\Lambda_0=2\Lambda\,,\label{setdue}
\end{equation}
where $M_{Pl}$ is the Planck mass and $\Lambda$ is the cosmological constant,
which is much smaller than the curvature at the inflation scale. By considering
$\lambda(0)$ of the order of the unit to avoid the $R^2$-correction at the
present epoch,
at the time of inflation one can put $\Lambda(t')=0$.

Let us assume that $R=R_\text{dS}$ describes the curvature of (de Sitter)
inflation.
Since it must be $R_0\ll R_\text{dS}\equiv 12H_\text{dS}^2$, where
$R_0=4\Lambda$, one has
\begin{equation}
\log\left[\frac{R_\text{dS}}{R_0}\right]=\log\left[H^2_{\text{dS}}\kappa_0^2\right]
-\log\left[\frac{\Lambda}{3}\kappa_0^2\right]\simeq
-\log\left[\frac{\Lambda}{3}\kappa_0^2\right]\,.\label{approx}
\end{equation}
Thus, from (\ref{tprime}) we get
\begin{equation}
t'\simeq -t_0'\log\left[\frac{\Lambda}{3}\kappa_0^2\right]\,,\quad 1\ll
t'\,,\label{tprimeapprox}
\end{equation}
namely $t'$ expresses the rate of the curvature of the current universe with
respect to the Planck mass on logaritm scale: this approximation is valid as
soon as $R_\text{dS}$ is near to $M_{Pl}^2$ during inflation, where ``near'' is
understood as ``with respect to the cosmological constant scale''.
In fact, the solution of Eq.~(\ref{dSeq}) depends on the value of today
$\lambda(0)$, which fixes the bound of inflation. From (\ref{dSeq}), we derive
the following solution,
\begin{equation}
H_\text{dS}^2\kappa_0^2\simeq
\frac{0.0146}{t_0'(\lambda(0)t')^{0.77}}\equiv
\frac{0.0146}{t_0'^{1.77}(\lambda(0))^{0.77}}\frac{1}{\left[-\log\left[\frac{\Lambda}{3}\kappa_0^2\right]\right]^{0.77}}\,,\label{setH}
\end{equation}
where we have taken into account that $1\ll t'$.
If we use the recent cosmological data~\cite{WMAP} for the evaluation of
$\Lambda$ in Planck units (see also Ref.~\cite{Barrow}),
\begin{equation}
\Lambda\kappa_0^2\simeq 1.7\times 10^{-121}\,,\label{setLambda}
\end{equation}
and we set for simplicity $t_0'=1$, we finally obtain
\begin{equation}
H_\text{dS}^2\kappa_0^2\simeq \frac{19\times
10^{-5}}{\lambda(0)^{0.77}}\,.\label{ex}
\end{equation}
For example, for $\lambda(0)=1$, we have
\begin{equation}
-\frac{\omega_2}{3\lambda(0)}\left(4\Lambda\kappa_0^2\right)R\simeq
4.53\times10^{-123}R\ll R\,,\quad
\frac{1.7\times10^{-121}}{3}M_{Pl}^2\simeq \left(\frac{\Lambda}{3}\right)\ll
H_\text{dS}^2\simeq 3.8\times 10^{-6}M_{Pl}^2\,.
\end{equation}
The first condition guarantees that at the present epoch the
$R_0^2$-contribution to the action (\ref{action}) is negligible with respect to
the Hilbert-Einstein term $R_0/\kappa_0^2$, where $R_0=4\Lambda$. The second
condition shows that
    de Sitter solution of inflation takes place at very high curvature near to
the
Planck scale, such that the approximation (\ref{approx}) is well satisfied.
We also note that during inflation
\begin{equation}
\frac{R}{\kappa^2(t')}\simeq 1.6\times 10^{-9}
M_{Pl}^4\ll-\frac{\omega(t')}{3\lambda(t')}R^2\simeq 5.1\times
10^{-8}M_{Pl}^4\,,
\end{equation}
and the second term in (\ref{action}) is dominant with respect to the
Hilbert-Einstein contribution at the early universe, thanks to the fact that
the running constant $\kappa^2(t')$ increases back into the past.

\section{Dynamics of inflation}

In this section, we would like to analyze the behaviour of the model
(\ref{action}) at high curvature, when the de Sitter solution describing
inflation (\ref{setH}) takes place. First of all, in order to have the exit
from inflation, one must show that the solution is unstable. Hence,
we can try to describe the inflation in terms of $e$-folds number and slow-roll
parameters.

\subsection{Instability of  de Sitter universe}

Let us consider the following form of Hubble parameter which is used in
Eq.~(\ref{sistemone}),
\begin{equation}
H=H_\text{dS}+\delta H(t)\,,\quad |\delta H(t)|\ll1,\label{approxH}
\end{equation}
where $\delta H(t)$ is the perturbation with respect  to de Sitter
    inflation. By making use of Eq.~(\ref{dSeq}) and
(\ref{setuno})--(\ref{setdue}) with $c_0\,,\Lambda=0$ in Eq.~(\ref{sistemone}),
and by multiplying it by $\kappa_0^3$, one has at the first order in $\delta
H(t)\equiv \delta H$,
\begin{eqnarray}
&&\hspace{-1cm}0=(\kappa_0\dot\delta H)\left[t_0'\left(
\left(H_{\text{dS}}\kappa_0\right)^2
\left(34.344 -\frac{0.913
      t_0'}{t'}\right)+\frac{0.001 t'+0.003 t_0'}{t'^3
\left(H_{\text{dS}}\kappa_0\right)^2 (\lambda(0)
      t')^{1.54}}+\frac{0.346 t_0'-0.086 t'}{t'^2 (\lambda(0)
t')^{0.77}}\right)
+19.152 t'
      \left(H_{\text{dS}}\kappa_0\right)^2\right]
\nonumber\\&&
\hspace{1cm}+\frac{(\kappa_0^2 \ddot\delta H )}{t'^3
\left(H_{\text{dS}}\kappa_0\right)^3
}\left[\phantom{\frac{0}{0}}t'^2
\left(H_{\text{dS}}\kappa_0\right)^4
      \left(6.384 t'^2+t_0' (11.448 t'-0.228 t_0')\right)
\right.
\nonumber\\
&&\left.
\hspace{1cm}
      -\frac{0.043 t'^2 t_0' \left(H_{\text{dS}}\kappa_0\right)^2}{(\lambda(0)
t')^{0.77}}+ \frac{0.001 t_0'^2}{(\lambda(0) t')^{1.54}}+\frac{0.087 t'
      t_0'^2 \left(H_{\text{dS}}\kappa_0\right)^2 }{(\lambda(0) t')^{0.77}}
+\frac{2\times 10^{-4} t' t_0' }{(\lambda(0)
      t')^{1.54}}\phantom{\frac{0}{0}}\right]\nonumber\\&&
\hspace{1cm}+\left(H_{\text{dS}}\kappa_0\right)
\delta H
      \left[\frac{0.223}{(\lambda(0) t')^{0.77}}+\frac{0.172 \lambda(0)
      t_0'}{(\lambda(0) t')^{1.77}}-30.528 t_0'
\left(H_{\text{dS}}\kappa_0\right)^2\right]\,.\label{sistpert}
\end{eqnarray}
If we assume
\begin{equation}
1\ll(H_\text{dS}\kappa_0)^2 t'^{2.27}\,,\label{condHtprime}
\end{equation}
the above expression is simplified as
\begin{equation}
D_0\delta H+
t'[19.152 (H_\text{dS}\kappa_0)(\kappa_0\dot \delta
H)+6.384(\kappa_0^2\ddot\delta H)]\simeq 0\,,\label{stabeq}
\end{equation}
where
\begin{equation}
D_0=
      \left(\frac{0.223}{(\lambda(0) t')^{0.77}}-30.528 t_0'
\left(H_{\text{dS}}\kappa_0\right)^2\right)\,.\label{D0}
\end{equation}
Thus, the solution of the equation reads
\begin{equation}
\delta H= h_\pm\exp\left[A_\pm t\right]\,,
\quad
A_\pm=
\left[\frac{H_{\text{dS}}}{2}\left(-3\pm\sqrt{9-\frac{0.627
D_0}{(H_\text{dS}\kappa_0)^2t'}}\right)\right]\,,
\quad
|h_{\pm}|\ll 1\,,\label{solpert}
\end{equation}
where $h_\pm$ are the integration constants corresponding to  plus and minus
signs inside $A_\pm$.
By choosing the sign plus in (\ref{solpert}), the solution is unstable under
the condition
\begin{equation}
D_0<0\,.\label{D}
\end{equation}
We would like to note that if we ignore the contribution from $\delta H$ in
(\ref{stabeq}),
we get
\begin{equation}
-\frac{\omega}{3\lambda}\left[(-216 H_{\text{dS}}^2)\dot\delta
H(t)-72\ddot\delta H(t)\right]\simeq 0\,,\label{HpertR2}
\end{equation}
which is the equation for perturbation around the de Sitter solution in pure
$R^2$-theory with Lagrangian $\mathcal L= -(\omega/(3\lambda))R^2$,
$\omega/3\lambda$ being constant. From this equation is not possible to know if
the solution is stable or not, since $\delta H$ mainly goes like $\delta H\sim
\text{const}$ in the time and even a
small contribution from the coefficient in front of $\delta H(t)$ could make
the solution unstable, such that a further analysis is required. In particular,
the fact that the coefficient in front to $R^2$ is not a constant contributes
to the instability of the solution, since for the Lagrangian $\mathcal L=
-(\omega(t')/(3\lambda(t')))R^2$ we get the equation
\begin{equation}
-\frac{\omega}{3\lambda}\left[(-216 H_{\text{dS}}^2)\dot\delta H-72\ddot\delta
H\right]
+(24
H_\text{dS})^2(6H_\text{dS})^3\frac{d}{dR}\left(\frac{\omega(t')}{3\lambda(t')}\right)\delta
H\simeq 0\,,\label{deltaHR2}
\end{equation}
where we have omitted the additional contributions to $\dot\delta H\,,\ddot
\delta H$. The term related to $\delta H$ corresponds to the last term of $D_0$
in (\ref{D0}), and, if it is dominant, it makes the solution
(\ref{solpert}) unstable.

Let us discuss the conditions (\ref{condHtprime}) and (\ref{D}). If
\begin{equation}
\frac{0.007}{t_0(\lambda(0) t')^{0.77}}<\left(H_\text{dS}\kappa_0\right)^2\,,
\end{equation}
both of the conditions are well satisfied and by taking into account de Sitter
solution (\ref{setH}) we see that this formula holds always true and it is
independendent on the bound of inflation encoded in $\lambda(0)$! It means,
that  de Sitter solution  is unstable with
\begin{equation}
D_0\simeq-\frac{0.223}{(\lambda(0)t')^{0.77}}\,,\label{D0one}
\end{equation}
where we have used (\ref{setH}). Moreover,
\begin{equation}
A_+\simeq 0.796\frac{H_\text{dS}t_0'}{t'}\,,\quad A_-\simeq -3
H_\text{dS}\,,\label{Apiu}
\end{equation}
where  $D_0$ has been considered very small.
For example, by setting $H_{\text{dS}}\kappa_0$ with
(\ref{setH})--(\ref{setLambda}) and by putting $t_0'=1$ and $\lambda(0)=1$, one
derives
\begin{equation}
\delta H=h_{-}\,\text{e}^{-5833 \times 10^{-6} M_{Pl} t}+
h_{+}\,\text{e}^{5.54 \times 10^{-6} M_{Pl} t}\,.\label{expert}
\end{equation}
During inflation, as soon as $t\ll 1/A_+$, avoiding the contribution of $h_{-}$
which quickly disappears, one may estimate
\begin{equation}
\delta H\simeq h_+\,,\quad\dot \delta H\simeq h_+A_+\,,\quad \ddot \delta
H\simeq h_+ A_+^2\,,
\label{apap}
\end{equation}
where $A_+$ is the instability parameter.
The duration of inflation $\Delta t$ is of the order of magnitude
\begin{equation}
\Delta t\sim\frac{1}{A_+}\,,\label{Deltaorder}
\end{equation}
but may continue after the linear approximation of the perturbation.
In the case of (\ref{expert}) one has
\begin{equation}
\Delta t\sim\frac{18\times 10^{4}}{M_{Pl}}\,. \label{exDelta}
\end{equation}
The inflation solves the problems of initial conditions of the Friedmann
universe (horizon and velocities problems), if $\dot a_\mathrm{i}/\dot a_0<
10^{-5}$, where $\dot a_\mathrm{i}\,,\dot a_0$ are the time derivatives of the
scale factor at the Big Bang and today, respectively, and $10^{-5}$ is the
estimated value of the inhomogeneity (anisotropy) in our universe. Since at
decelerating universe $\dot a(t)$ decreases by a factor $10^{28}$, it is
required that $\dot a_\mathrm{i}/\dot a_\mathrm{f}<10^{-33}$, with
$a_\mathrm{i}$ the scale factor at the beginning of inflation and
$a_\mathrm{f}$ the scale factor at the end of inflation. If inflation is
governed by a (quasi) de Sitter solution where
$a(t)=\exp\left(H_\text{dS}t\right)$, we introduce the number of $e$-folds $N$
as
\begin{equation}
N=\ln \left(\frac{a_\mathrm{f}}{a_\mathrm{i}}\right)\equiv\int^{t_f}_{t_i} H(t)
dt\,,
\end{equation}
and inflation is viable if $N>76$, but the spectrum of fluctuations of CMB say that
it is enough $\mathcal N\simeq 55$ to have thermalization of observable universe. In
our case,
\begin{equation}
N\simeq H_\text{dS}\Delta t\sim
\frac{H_\text{dS}}{A_+}
\simeq 1.26\left(\frac{t'}{t_0'}\right)\,,\label{N}
\end{equation}
due to the fact that the Hubble parameter is almost a constant during
inflation. In order to obtain a viable inflation it must be
\begin{equation}
61<\left(\frac{t'}{t_0'}\right)\,.
\end{equation}
It means, from (\ref{tprime}) and (\ref{setLambda}),
\begin{equation}
3.1 R_0\times 10^{26}<R\,,
\end{equation}
and this condition is always satisfied for realistic inflation.
For the case of  (\ref{ex}), where the Hubble parameter during inflation is 117
times larger than today and whose
duration of inflation is given by (\ref{exDelta}), we get
\begin{equation}
N\sim 339\,,\label{Nex}
\end{equation}
and it is guaranteed the thermalization of a portion of universe much larger
with respect to the observed one.

It is clear that a large $e$-folds number, which corresponds to a huge amount of
inflation, may be related to the fact that the universe remains extremely close to
the de Sitter space-time during inflation.
In fact, even if, without additional data about the decay of the primordial
accelerated expansion (the so called ``false vacuum''), we cannot pose any upper
limit to the $e$-folds number and we could expect that the homogeneity and isotropy
continue for some distance beyond our observable universe, the primordial
perturbations at the end of inflation depend on the $e$-folds.
As a consequence, as we will see in the next subsection, a
large $e$-folds could generate wrong predictions for the spectral index.
In the last part of the work we will find how it is possible to make
inflation shorter according with a correct prediction of such index.

\subsection{Slow-roll parameters and spectral index\label{ss}}

During the inflation the Hubble parameter must slowly decrease and the
following approximations must be meet,
\begin{equation}
|\frac{\dot H}{H^2}|\ll 1\,,\quad |\frac{\ddot H}{H\dot H}|\ll
1\,.\label{slowrollcondition}
\end{equation}
Thus, one introduces the slow-roll parameters
\begin{equation}
\epsilon=-\frac{\dot H}{H^2}\,,\quad\eta=-\frac{\dot H}{H^2}-\frac{\ddot H}{2 H\dot
H}\equiv2\epsilon-\frac{1}{2\epsilon H}\dot\epsilon\,,\label{slowrollpar}
\end{equation}
whose magnitude must be small during inflation and $\dot H$ is assumed to be
negative. In particular, since the acceleration is expressed as
\begin{equation}
\frac{\ddot a}{a}=\dot H+H^2\,,
\end{equation}
we see that the universe expands in accelerated way as soon as $\epsilon<1$.
By integrating the formula for the (positive and almost constant) $\epsilon$
parameter in (\ref{slowrollpar}) we also get
\begin{equation}
H(t)=\frac{1}{\epsilon (t_\text{dS}+t)}\,,\quad
t_\text{dS}\simeq\frac{1}{\epsilon H_\text{dS}}\,,\label{Hepsilon}
\end{equation}
where $t_\text{dS}$ is a positive time parameter and when the time increases
the Hubble parameter decreases. In the limit $t/t_\text{dS}\ll 1$, one has
\begin{equation}
H(t)\simeq H_\text{dS}-H^2_\text{dS}\epsilon t\,,\label{Hepsilon}
\end{equation}
and by taking into account (\ref{apap}) we get
\begin{equation}
\epsilon\simeq\frac{(-h_+) A_+
}{(H_\text{dS})^2}=0.796272\left(\frac{t_0'}{t'}\right)\frac{(-h_{+})}{H_\text{dS}}\,,\label{epsilonrel}
\end{equation}
where $h_+<0$ and $A_+$ is given by (\ref{Apiu}).
This relation is consistent with a direct evaluation of the slow-roll parameter
$\epsilon$ (\ref{slowrollpar}) in the slow-roll limit (\ref{slowrollcondition})
of the equation of motion
(\ref{sistemone}),
\begin{eqnarray}
0&=&
2 \lambda ^2 \epsilon\left[480 H^4 \kappa^4 \omega ^3 (4 \omega +3) (2 \omega
      (100 \omega +549)+25)+2 \kappa ^4 \lambda  \Lambda  \omega  (4 (355-1616
\omega )
      \omega +45)
\right.
\nonumber\\&&
\left.
-15 \lambda ^2 (\omega  (2 \omega  (4 \omega  (50 \omega
      +97)-25)-71)-5)\right]
+720 \kappa ^2 \omega ^3 \left(72 H^4 \kappa ^2 \omega
      \epsilon +6 H^2 \lambda -\kappa ^2 \lambda  \Lambda \right)
\nonumber\\&&
+15 \lambda  \omega
      \left(-480 H^4 \kappa ^4 \omega ^2 (4 \omega  (2 \omega +3)+1) (8 \epsilon
+1)+4
      \kappa ^2 \lambda  \omega  \left(3 H^2 (4 \omega -3) (10 \omega +1) (7
\epsilon
      +2)+2 \kappa ^2 \Lambda  (28 \omega +1)\right)+
\right.
\nonumber\\&&
\left.
\lambda ^2 \left(60 \omega
      ^2+3\right)\right)\,.
\end{eqnarray}
    By using (\ref{setuno})--(\ref{setdue}) with $c_0=\Lambda=0$, one obtains the
solution
\begin{eqnarray}
\hspace{-0.5cm}\epsilon&\simeq&
\frac{-\frac{3\times 10^{-4} t'_0}{t'^2 (\lambda(0) t')^{1.54}}-\frac{0.086
      \lambda(0) t'_0 \left(H\kappa_0\right)^2}{(\lambda(0)
      t')^{1.77}}-\frac{0.112 \left(H\kappa_0\right)^2}{(\lambda(0)
t')^{0.77}}+7.632 t_0
      \left(H\kappa_0\right)^4}{-\frac{0.003 (t_0')^2}{t'^3 (\lambda(0)
      t')^{1.54}}+\frac{0.913 (t_0')^2
\left(H\kappa_0\right)^4}{t'}+t'_0\left(H\kappa_0\right)^2
      \left(\frac{0.301 \lambda(0)}{(\lambda(0) t')^{1.77}}-61.056
      \left(H\kappa_0\right)^2\right)-19.152 t' \left(H\kappa_0\right)^4}\,,
\end{eqnarray}
and under the condition (\ref{condHtprime}) we derive
\begin{equation}
\epsilon\simeq\frac{0.006}{t'(\lambda(0)t')^{0.77}(H\kappa_0)^2}-\frac{0.398}{t'}\,.
\end{equation}
By expanding $H(t)$ around  de Sitter solution (\ref{setH}) we finally get
\begin{equation}
\epsilon\simeq\frac{-2(0.006)}{t'(\lambda(0)t')^{0.77}(H_\text{dS}\kappa_0)^3}\kappa_0\delta
H=\frac{0.012}{t'(\lambda(0)t')^{0.77}(H_\text{dS}\kappa_0)}\frac{\epsilon}{\kappa_0
A_+}\,,
\end{equation}
where Eqs.~(\ref{apap}) and (\ref{epsilonrel}) are considered: the equation is
well satisfied by using (\ref{setH}) again and (\ref{Apiu}).
Thus, the $\epsilon$ slow-roll parameter is related to the (initial) amplitude
of perturbation
and by using (\ref{N}) one may estimate
\begin{equation}
\epsilon\simeq\frac{(-h_+) A_+ }{(H_\text{dS})^2}
\sim\frac{(-h_+)}{(H_\text{dS}) N}
\,.\label{epsilon}
\end{equation}
Moreover, for the $\eta$ slow-roll parameter in (\ref{slowrollpar}) with
(\ref{apap}) one has
\begin{equation}
\eta\simeq-\frac{A_+}{2H_\text{dS}}\simeq-\frac{0.398
t_0'}{t'}\sim\frac{1}{2N}\,.\label{eta}
\end{equation}
Both of the paramter $\epsilon\,, |\eta|$ (\ref{epsilon})--(\ref{eta}) are very
small during inflation and the slow-roll approximations
(\ref{slowrollcondition}) hold true.
We also note that, since $|h_+|\ll H_\text{dS}$,
\begin{equation}
\epsilon\ll|\eta|\,,\label{ll}
\end{equation}
like in other scalar tensor theories for inflation, where usually $\epsilon\sim 1/N^2$,
as in (\ref{epsilon}) if we consider $(-h_+)/H_\text{dS}\sim 1/N$.

Given the slow-roll parameters, one can evaluate the universe anisotropy coming
from inflation by introducing the spectral indexes.
To be specific, the amplitude of the primordial scalar power spectrum reads
\begin{equation}
\Delta_{\mathcal R}^2=\frac{\kappa^2 H^2}{8\pi^2\epsilon}\,,\label{spectrum}
\end{equation}
and for slow-roll inflation the spectral index $n_s$ and the tensor-to-scalar
ratio are given by
\begin{equation}
n_s=1-4\eta\,,\quad r=48\epsilon^2\,,\label{indexes}
\end{equation}
where we use the results for modified gravity~\cite{corea}.
The last Planck data~\cite{WMAP} constrain these quantities as
\begin{equation}
n_s=0.9603\pm0.0073\,,\quad r<0.11\,.\label{data}
\end{equation}
For our model one has the scalar power spectrum
\begin{equation}
\Delta_\mathcal R\simeq
1.25585\left(H_\text{dS}\kappa_0\right)^{3}
\left(\frac{t'}{t_0'}\right)(-\kappa^2_0 h_+)^{-1}\,,
\end{equation}
and the spectral index and the tensor-to-scalar ratio,
\begin{equation}
n_s=1-\frac{2 A_+}{H_\text{dS}}\sim 1-\frac{2}{N}
\,,
\quad
r=\frac{48A_+^2}{H_\text{dS}^2}\frac{(-h_+)^2}{H_\text{dS}^2}\ll\frac{1}{N}
\,,
\end{equation}
where we have used (\ref{ll}).
We see that the tensor-to-scalar
ratio can satisfy the Planck results, being the $e$-folds of realistic
inflation quite large.
On the other side,
in order to find the spectral index $n_s$ in agreement with the Planck data
(\ref{data}), we must require
\begin{equation}
21<\frac{2 A_+}{H_\text{dS}}\left(=2.5117\left(\frac{t'}{t_0'}\right)\right)<31\,,
\end{equation}
Since $A_+/H_\text{dS}$ depends on the ratio between the curvature of the
universe at the time of inflation and the curvature of today universe, it
results particulary high and does not satisfy this condition, contributing to
render near to one the spectral index $n_s$ of the model. For example, in the
case of (\ref{ex}) where the Hubble parameter during inflation is 117 times
larger than today and the $e$-folds $N\sim
H_\text{dS}/(A_+)\simeq 339$ as in (\ref{Nex}),
\begin{equation}
n_s\simeq 0.994
\,,
\quad
r\simeq 0.0004\frac{(-h_+)^2}{H_\text{dS}^2}
\,.
\end{equation}
Since $(-h_+/H_\text{dS})\ll 1$, the tensor-to-scalar ratio is much smaller than
$0.11$, but the spectral index does not satisfy the Planck data. This should be
compared with analysis of inflationary parameters for general $F(R)$-theory in
fluid-like presentation \cite{Bamba} which maybe consistent with Planck
data.

The large $e$-folds number and the $n_s$ spectral index too close to one are
consequences of the small value of $A_+$ (\ref{Apiu}), which depends on
$d(\omega(t')/3\lambda(t'))/d t'$, as we explained under (\ref{deltaHR2}). In
particulary, the fact that $d(\omega(t')/3\lambda(t'))/d t'=-\beta_3/3$, where
$\beta_3$ is given in (\ref{betabeta}), such that $\beta_3\ll 1$, makes this
term too small compared with the coefficients in front of $\ddot\delta H(t),
\dot\delta H(t)$ in the equation for perturbation (\ref{HpertR2}). In the next
section, we  suggest a possible solution of the problem returning to the
general action (\ref{action}) with the Gauss-Bonnet and $\Box R$ terms which
have been omitted in the above study.

\section{ The account of Gauss-Bonnet and $\Box R$ terms and spectral
index\label{GB}}

As it was mentioned in second section,  to construct the  Lagrangian of
higher-derivative gravity, also the Gauss-Bonnet and the $\Box R$ terms must be
taken into account. They may give a non-zero contribution to the dynamical
equations
if the coefficients in front of them are not constant but depend on the
curvature. This is precisely what happens when one solves RG equation and gets
RG improved effective action. In the first part of this work we did not
consider such contributions. Let us  analyze their role on the dynamics of the
inflation induced by higher-derivative quantum gravity.
Let us consider the following additional piece to the action (\ref{action}),
\begin{equation}
I_{G\,,\Box R}=-\int_\mathcal M d^4 x\sqrt{-g} \left[\gamma(t') G-\zeta(t')\Box
R\right]\,,\label{new}
\end{equation}
where $G$ is given by (\ref{Gauss}) and $\gamma(t')\,,\zeta(t')$ are effective
coupling  constants depending on $t'$ (\ref{tprime}) and therefore on $R$.
We assume
\begin{equation}
\gamma(t')=\gamma_{0}(1+c_1
t')\,,\quad\zeta(t')=\zeta_0(1+c_2 t')\,,\label{gammazetalimit}
\end{equation}
where $\gamma_0\,,\zeta_0$ are generic constants and $c_{1,2}$ are numerical
coefficients whose explicit values are not necessary in the below
analysis.
As it is explained in review \cite{B} this is result of one-loop
quantum calculation of these terms (vacuum polarization).
For recent discussion of contribution of GB term in higher-derivative
gravity, see Ref.~\cite{GBdisc}.
Actually, the calculation of surface terms may be done in less/more
than four dimensions, with subsequent
dimensional continuation.

Hence,
when $t'\ll 1$, at the low curvature limit, $\gamma(t')\,,\zeta(t')$ tend
to
constants, the derivatives do not diverge and (\ref{new}) turns out to be zero:
on the other side, when $1\ll t'$, at the high curvature limit, they give
a
significative contribution to the dynamical equations of motion.
The Gauss-Bonnet represents a new curvature invariant. On FRW metric it
(\ref{metric}) reads
\begin{equation}
G=\frac{24\dot a^2}{a^3 N^5}\left(\ddot a N-\dot a \dot
N\right)\,.\label{GBgeneral}
\end{equation}
Adding to the Lagrangian (\ref{action2}) the piece (\ref{new}), we make an
integration by parts with respect to $\Box R$, where $\Box
R=(\sqrt{-g})^{-1}\partial_\mu (g^{\mu\nu}\sqrt{-g}\partial_\nu R)\equiv
-(\sqrt{-g})^{-1}\partial_t (\sqrt{-g}\partial_t R)$, and introduce a new
Lagrangian multiplier $\sigma$ for the Gauss-Bonnet term~\cite{Monica},
such that
\begin{equation}
I_{G\,,\Box R}=-\int_\mathcal M d^4 x\sqrt{-g} \left[\gamma(t')
G+\sigma\left[G-\frac{24\dot a^2}{a^3 N^5}\left(\ddot a N-\dot a \dot
N\right)\right]-\left(\frac{d\zeta}{d t'}\frac{d t'}{d A}\dot
A^2\right)\right]\,,\quad \sigma=-\gamma(t')\,,\label{lll}
\end{equation}
Here the second expression has been derived from the variation with
respect to $G$
and $A\equiv A(N,\dot N, a, \dot a)$ is the explicit form of the Ricci scalar
as a function of the metric (\ref{Ricci}),
\begin{equation}
A(N,\dot N, a, \dot a)=\frac{1}{N^2}\left[6\left(\frac{\dot
a}{a}\right)^2+6\left(\frac{\ddot a}{a}\right)-6\left(\frac{\dot
N}{N}\right)\left(\frac{\dot a}{a}\right)\right]\,.
\end{equation}
Thus, $\Delta(t')$ in (\ref{xi}) reads
\begin{equation}
\Delta(t')=\left[\frac{R}{(\kappa^2(t'))^2}\frac{d\kappa^2(t')}{d
t'}+R^2\frac{d}{d
t'}\left(\frac{\omega(t')}{3\lambda(t')}\right)+\frac{d\Lambda(t')}{d
t'}+\frac{d\gamma (t')}{d t'}G\right]\,, \label{newDeltatprime}
\end{equation}
and the additional piece to the Lagrangian (\ref{lagrangian}) results to be
\begin{equation}
\mathcal L_{G\,,\Box R}(N, \dot N, \ddot N, a,\dot a, \ddot a, R, \dot R)=6\dot
a\left(\frac{a^2}{N}\right)\left[\frac{d\gamma(t')}{d t'}\frac{d t'}{d R}\dot
G\right]+\frac{8\dot a^3}{N^3}\frac{d\gamma(t')}{d t'}\frac{d t'}{d R}\dot
R+(Na^3)\left(\frac{d\zeta}{d t'}\frac{d t'}{d A}\dot A^2\right)\,,
\end{equation}
where the first piece comes from the integration by parts of the second
derivative metric functions of the Ricci scalar, the second term comes from the
ones of the Gauss-Bonnet and the last piece corresponds to $\Box R$
-term. Note that now the Lagrangian depends on the higher derivatives of the
metric due to the introduction of $\dot A^2$.
Equation (\ref{NN2}), in the gauge $N=1$, is derived as
\begin{eqnarray}
0&=&
-\Lambda(t')+\frac{6 H^2}{\kappa^2(t')}-\frac{6
H}{(\kappa^2(t'))^2}\frac{d\kappa^2(t')}{d t'}\left(\frac{t_0' \dot
R}{R}\right)
+\frac{\omega(t')}{3\lambda(t')}
\left[6R\dot H
-12H\dot R\right]
-12 H\frac{d}{d t'}\left(\frac{\omega(t')}{3\lambda(t')}\right)\left(\dot R
t_0'\right)\nonumber\\&&
+6\left(H^2+\dot H\right)\Delta(t')\frac{t_0'}{R}-6 H
\left[\frac{d\Delta(t')}{d
t'}\left(\frac{t_0'}{R}\right)^2-\Delta(t')\frac{t_0'}{R^2}\right]\dot R
-24 H^3\frac{d\gamma(t')}{d t'}\frac{t'_0\dot R}{R}-6
H\left[\frac{d\gamma(t')}{d t'}\frac{t_0'\dot G}{R}\right]\nonumber\\&&
-3\mathcal A\dot R^2
-2\mathcal B\dot R^2 R
+6\frac{d}{d t}\left[2\mathcal A\left(4H^2+3\dot H\right)\dot R+\mathcal B
H\dot R^2\right]
+18 H\left[2\mathcal A\left(4H^2+3\dot H\right)\dot R+\mathcal B H\dot
R^2\right]\nonumber\\
&&-36\left(3H^2+\dot H\right)\mathcal A\,H\dot R-72H\frac{d}{d t}\left(\mathcal
A H \dot R\right)-12\frac{d^2}{d t^2}
\left(\mathcal A H \dot R\right)
\,,
\label{NN2new}
\end{eqnarray}
where
\begin{equation}
\mathcal A=\left(\frac{d\zeta(t')}{d t'}\frac{t'_0}{R}\right)\,,\quad
\mathcal B=\left[\frac{d^2\zeta(t')}{d
t'^2}\left(\frac{t'_0}{R}\right)^2-\frac{d\zeta(t')}{d
t'}\frac{t'_0}{R^2}\right]\,,
\end{equation}
and the Ricci scalar $R$ is given by (\ref{R2}).
The derivative of the Lagrangian with respect to the Gauss-Bonnet leads to the
Ricci scalar in (\ref{Ricci}), and the derivative with respect to the
Ricci scalar
leads to the Gauss-Bonnet one in (\ref{GBgeneral}), which reads in the
gauge $N=1$,
\begin{equation}
G=24H^2\left(H^2+\dot H\right)\,.
\end{equation}
On  de Sitter solution $R_\text{dS}=12 H_\text{dS}^2$, $G_\text{dS}=24
H_\text{dS}^4$, $H_\text{dS}$ being constant, equation (\ref{dSeq}) is
corrected as
\begin{eqnarray}
0&=&
\frac{6 H^2}{\kappa ^2}-\frac{t_0'}{48 (\kappa ^2)^2 \omega ^2} \left(480 H^4
(\kappa^2)^2 \omega ^2 (4 \omega  (2
      \omega +3)+1)+4 \kappa ^2 \lambda  \omega  \left(6 H^2 \left(-40 \omega
^2+26 \omega
      +3\right)-2 \kappa ^2 \Lambda  (28 \omega +1)\right)
\right.
\nonumber\\&&
\left.
-3 \lambda ^2 \left(20 \omega
      ^2+1\right)\right)-\Lambda+12H^4\frac{d\gamma}{d t'}t_0'\,,
\label{dSeq2}
\end{eqnarray}
where the functions $\lambda\,,\omega\,,\kappa^2\,,\Lambda$ and $\gamma\,,
d\gamma/d t'$ are constants in the time. By using
(\ref{setuno})--(\ref{setdue}) with $c_0=\Lambda=0$, and $1\ll t'$, we obtain
the
solution
\begin{equation}
H_\text{dS}^2\kappa_0^2\simeq
\frac{322.762}{\left(22085.2-34725.2(d\gamma/dt')\right)t_0'(\lambda(0)t')^{0.77}}\,,\quad\frac{d\gamma}{d
t'}<0\,,\label{dS2}
\end{equation}
where   $|d\gamma/d t'|\ll t'^2$ is used and we require
that such a derivative is negative ($\gamma_0c_1<0$ in (\ref{gammazetalimit})).
Thus, given the form of $\gamma_{t'}(t')$,  de Sitter solution
depends on the current value of $\lambda(t'=0)=\lambda(0)$.
Obviously, the $\Box R$-term does not give any contribution to the de Sitter
solution.
By using again the parametrization (\ref{setuno})--(\ref{setdue}) with
$c_0=\Lambda=0$, and therefore by multiplying (\ref{NN2new}) by $\kappa_0^3$,
and
by perturbationg it with respect to  de Sitter solution (\ref{dS2}) as in
(\ref{approxH}), we get
\begin{eqnarray}
&&\frac{\kappa_0}{t'^3 (H_\text{dS}\kappa_0)^3}
    \left[\kappa_0 \ddot\delta H \left(t'^2 (H_\text{dS}\kappa_0)^4  \left(6.384
t'^2+t' t_0'(-18\gamma_{t'}(t') +18\zeta_{t'}(t')
-6\gamma_{t't'}(t') t_0'+11.448)
\right.\right.\right.
\nonumber\\
&&\left.
-0.228 t_0'^2\right)-0.043t'^2 t_0'\left(H_\text{dS}\kappa_0\right)^2
(\lambda(0) t')^{-0.77}+0.001 t_0'^2 (\lambda(0) t')^{-1.54}+0.087 t'
t_0'^2 \left(H_\text{dS}\kappa_0\right)^2
      (\lambda(0) t')^{-0.77}
\nonumber\\&&
\left.
+2\times 10^{-4} t' t_0'(\lambda(0)
t')^{-1.54}\right)+\left(H_\text{dS}\kappa_0\right)
      \dot\delta H \left(t'^2 \left(H_\text{dS}\kappa_0\right)^4 \left(19.152
t'^2+t' t_0'(-54\gamma_{t'}(t')
      +72 \zeta_{t'}(t')
\right.\right.
\nonumber\\&&
\left.\left.
-24 \gamma_{t't'}(t') t_0'+34.344)-0.913
      t_0'^2\right)-0.086 t'^2 t_0'\left(H_\text{dS}\kappa_0\right)^2
(\lambda(0) t')^{-0.77}+0.003
      t_0'^2 (\lambda(0) t')^{-1.54}
\right.
\nonumber\\&&
\left.\left.
+0.346 t' t_0'^2 \left(H_\text{dS}\kappa_0\right)^2 (\lambda(0)
      t')^{-0.77}+0.001 t' t_0'(\lambda(0)
t')^{-1.54}\right)\right]+\left(H_\text{dS}\kappa_0\right) \delta H
\left[\frac{0.223}{(\lambda(0)
      t')^{0.77}}
\right.
\nonumber\\&&
\left.
+t_0'\left(\frac{0.172 \lambda(0)}{(\lambda(0)
t')^{1.77}}+\left(H_\text{dS}\kappa_0\right)^2
      (48\gamma_{t'}(t') -30.528 )\right)\right]=0\,, \quad
H(t)=H_\text{dS}+\delta H(t)\,,\quad |\delta H(t)|\ll 1\,,
\end{eqnarray}
where we introduced the notation
\begin{equation}
\gamma_{t'}(t')\equiv\frac{d\gamma(t')}{d t'}\,,\quad\gamma_{t'
t'}(t')\equiv\frac{d^2\gamma(t')}{d t'^2}\,,\quad
\zeta_{t'}(t')\equiv\frac{d\zeta(t')}{d t'}\,.
\end{equation}
If one assumes (\ref{condHtprime}) and takes into account that
$|\gamma_{t'}(t')|\,,|\zeta_{t'}(t')|\ll t'$ and $|\gamma_{t' t'}(t')|\ll 1$,
this expression is
simplified as
\begin{equation}
\tilde D_0\delta H+
t'[19.152 (H_\text{dS}\kappa_0)(\kappa_0\dot \delta
H)+6.384(\kappa_0^2\ddot\delta H)]\simeq 0\,,\label{stabeq2}
\end{equation}
where
\begin{equation}
\tilde D_0=
      \left[\frac{0.223}{(\lambda(0) t')^{0.77}}-(30.528-48\gamma_{t'}(t'))
t_0' \left(H_{\text{dS}}\kappa_0\right)^2\right]\,.\label{D02}
\end{equation}
Thus, the solution of the above differential equation  reads
\begin{equation}
\delta H= h_\pm\exp\left[\tilde A_\pm t\right]\,,
\quad
\tilde A_\pm=
\left[\frac{H_{\text{dS}}}{2}\left(-3\pm\sqrt{9-\frac{0.627 \tilde
D_0}{(H_\text{dS}\kappa_0)^2t'}}\right)\right]\,,
\quad
|h_{\pm}|\ll 1\,,\label{solpert2}
\end{equation}
where $h_\pm$ are the integration constants corresponding to the
signs: plus and
minus inside $\tilde A_\pm$.
The solution is unstable if $\tilde D_0<0$, namely
\begin{equation}
\frac{0.223074}{(\lambda(0) t')^{0.77}}<
[30.528-48\gamma_{t'}(t')] t_0' \left(H_{\text{dS}}\kappa_0\right)^2\,,
\end{equation}
and, by using (\ref{dS2}), one sees that this inequality is always
satisfied
independently on the value of $\gamma_{t'}(t')$. As a consequence, also
(\ref{condHtprime}) that we have used to derive (\ref{stabeq2}) is verified and
it is interesting to note that $\tilde D_0$ evaluated with respect to  de
Sitter solution (\ref{dS2}) is equal to $D_0$ in (\ref{D0one}) evaluated
with respect to  de Sitter solution (\ref{setH}),
from which we can understand that  Gauss-Bonnet term contribution to the
stability of  de Sitter solution behaves like the one of a $R^2$-term (see
(\ref{HpertR2})--(\ref{deltaHR2}) and related comment).
By using (\ref{dS2}) one gets
\begin{equation}
\tilde A_+\simeq 36019\times
10^{-9}\frac{H_\text{dS}t_0'}{t'}\left(22085.2-34725.2\gamma_{t'}(t')\right)\,,\quad
\tilde A_-\simeq -3 H_\text{dS}\,,\label{Apiu}
\end{equation}
where  $\tilde D_0$ is taken to be small.
Thanks to the presence of the Gauss-Bonnet term in the action, the
instability
parameter $\tilde A_+$ can be increased with respect to the case considered
   before. Let us introduce our Ansatz (\ref{gammazetalimit}). We
obtain
\begin{equation}
H_\text{dS}^2\kappa_0^2\simeq
\frac{322.762}{\left[22085.2-34725.2\gamma_0c_1\right]t_0'(\lambda(0)t')^{0.77}}\,,\quad
\gamma_0c_1<0\,,\label{HHH}
\end{equation}
\begin{equation}
\tilde A_+\simeq 36019\times
10^{-9}\frac{H_\text{dS}t_0'}{t'}\left(22085.2-34725.2\gamma_0c_1\right)\,,\quad
\tilde A_-\simeq -3 H_\text{dS}\,.
\end{equation}
As a consequence, the instability parameter $\tilde A_+$ is larger than $A_+$
in the
absence of Gauss-Bonnet correction if $\gamma_0c_1$ is negative, namely, by
taking $0<c_1$ and $\gamma_0<0$, the Gauss-Bonnet contribution to the
action is positive (see (\ref{new})): the
analysis of inflation is similar to the previous case, but the $e$-folds and
therefore the spectral index $n_s$ are smaller.

To be specific, the $\eta$ slow-roll parameter (\ref{eta}) and the
spectral
index $n_s$ in (\ref{indexes})  read
\begin{equation}
\eta\simeq-\frac{18 \times 10^{-6}
t_0'\left(22085.2-34725.2\gamma_0 c_1\right)}{t'}
\,,\quad
n_s\simeq 1-
\frac{72038\times
10^{-9}t_0'\left(22085.2-34725.2\gamma_0c_1\right)}{t'}\,,\label{eta2}
\end{equation}
since we can still use (\ref{ll}).
The spectral index $n_s$ is consistent with Planck data (\ref{data}) if
\begin{equation}
450<\frac{t'_0}{t'}\left(22085.2-34725.2\gamma_0c_1\right)<653\,.\label{limitns}
\end{equation}
If we set $\lambda(0)=t_0=1$ and take (\ref{tprimeapprox}) together with
(\ref{setLambda}), we get from (\ref{limitns}),
\begin{equation}
-4.61<\gamma_0 c_1<-2.98\,.
\end{equation}
For example, for $c_1=1$ and $\gamma_0=-3$ we find
\begin{equation}
n_s\simeq 0.96740\,,
\end{equation}
which is in agreement with the Planck data (\ref{data}). The de Sitter solution
results to be $H_\text{dS}^2\simeq 3.17\times 10^{-7} M_{Pl}^2$, and inflation
takes place near to the Planck scale, such that (\ref{tprimeapprox}) is valid.
In this kind of model, as we noted in \S \ref{ss}, the e-folds $N\sim
2/(1-n_s)$, and in the present case we have $N\sim 60$: this is an order of
magnitude/lower bound of the $e$-folds which permits the thermalization of
observable universe
(the acceleration finishes when
$\epsilon=1$, and
therefore the exact amount of inflation depends on the initial amplitude
$|h_+|$ as in (\ref{epsilon})). Thanks to the Gauss-Bonnet contribution in the
action, we can see that the value of the $e$-folds has considerably decreased (see
for example (\ref{Nex})), rendering correct the prediction of the spectral index.
In the present
example, a viable inflation is obtained for $1\ll t'/t_0'$, which is
always true due to the large curvature scale of inflation.

We have demonstrated that the contribution from RG improved Gauss-Bonnet
term can modify the instability of
   de Sitter solution describing inflation given a viable spectral index. In
our derivation, we have taken into account also the $\Box R$ contribution, but,
due to the Ansatz (\ref{gammazetalimit}), it disappears. However, we furnished
the formalism to treat the Lagrangian (\ref{new}) with generic
coefficients: if they grow up in the early-time universe, they modify the
dynamics of inflation and can lead to a model compatible with the Planck
data.\\
\\
As a final result of the work, we are able to present the very general
quantum-corrected Lagrangian constructed with second degree corrections to
the Einstein gravity:
\begin{equation}
I=\int_\mathcal{M}d^4\sqrt{-g}\left[\frac{R}{\kappa^2(t')}-\frac{\omega(t')}{3\lambda(t')}R^2+\frac{1}{\lambda(t')}C^2-\gamma(t')
G+\zeta(t')\Box
R-\Lambda(t')\right]\,,\quad
t'=\frac{t'_0}{2}\log\left[\frac{R}{R_0}\right]^2\,,\label{acf}
\end{equation}
where $t_0$ is a number and $R_0=4\Lambda$ is the curvature of today
universe, $\Lambda$ being the cosmological constant.
The one-loop running coupling constants
$\lambda(t')\,,\omega(t')\,,\kappa^2(t')\,,\Lambda(t')\,,\gamma(t')$ and
$\zeta(t')$ are found from higher-derivative quantum gravity. They can be
written as
\begin{equation}
\lambda(t')=\frac{\lambda(0)}{(1+\lambda(0)(133/10) t')}\,,\quad
\omega(t')=\omega_{1}\,,\quad
\kappa^2(t')=\kappa_0^2(1+\lambda(0)(133/10) t')^{0.77}\,,\quad
\Lambda(t')=\frac{\Lambda_0}{(1+\lambda(0)(133/10) t')^{0.55}}\,,
\end{equation}
with $\omega_1=-0.02$, $\kappa_0^2=16\pi/M_{Pl}^2$, $\Lambda_0=2\Lambda$.
The expressions for $\omega(t')\,,\kappa^2(t')$ and $\Lambda(t')$ are
derived by investigating the asymptotic behaviour of the running constants at
high curvature. However, the
derivatives of the coupling constants obey to a set of RG equations that
we
have taken into account in our analysis.
The
form of $\gamma(t')$ and $\zeta(t')$ is given by
\begin{equation}
\gamma(t')=\gamma_0(1+c_1 t')\,,\quad\zeta(t')=\zeta_0(1+c_2 t')\,,\quad
c_1\gamma_0<0\,,\label{An}
\end{equation}
$\gamma_0\,,\zeta_0$ and $c_{1,2}$ constants.
Finally, $\lambda(0)$ is a number related to the bound of inflation. At small
curvature ($t'\ll 1$), the action (\ref{acf}) reads
\begin{equation}
I=\int_\mathcal{M}d^4\sqrt{-g}\left[\frac{R}{\kappa_0}+\frac{0.02}{\lambda(0)}R^2+\frac{1}{\lambda(0)}C^2-2\Lambda\right]\,,\quad
t'=\frac{t'_0}{2}\log\left[\frac{R}{R_0}\right]^2\,,
\end{equation}
and the contributions of Gauss-Bonnet and $\Box R$-terms disappear when
the
coefficients become constant.

Inflation is described at high curvature for $1\ll t'$, near to the Planck
mass. The model possesses a de Sitter solution which depends on
$\lambda(0)$.
This
solution is always unstable and the model exits from inflation.
It is possible to calculate the behaviour of perturbations and show that the
slow-roll conditions of inflation are satisfied with the $\epsilon$ slow-roll
parameter much smaller than the $\eta$ slow-roll parameter. The amount of
inflation ($e$-folds) is sufficiently large, the tensor-to-scalar ratio $r$ is
very close to zero and, due to the contribution of the RG improved
Gauss-Bonnet term in the
action, the spectral index $n_s$ satisfies the Planck data. The RG
improved $\Box
R$-term
does not play any important role in the dynamics
of inflation.

After inflation, the reheating process with the particle production must take
place to recover the FRW universe. These processes occur when the
curvature (Ricci scalar) oscillates and eventually in the presence of the
interaction
between the gravity and matter quantum fields. At the end of
inflation $t'\rightarrow 0$ and the model turns out to be a quadratic
correction
$R^2$ of Einstein's gravity (on FRW metric the square of Weyl tensor gives a
zero contribution): this model has been well-investigated in the
literature and it has been demonstrated that it is compatible with the
reheating scenario.

\section{Discussion}

In this work we  investigated the inflationary universe taking into
account  quantum gravity effects in frames of RG improved effective
   action of higher-derivative quantum gravity. The effective
coupling constants  in higher-derivative quantum gravity obey to a set
of one-loop RG equations found in Refs.~\cite{Fradkin} and may show the
asymptotically-free behaviour. These one-loop RG equations which define
the effective coupling constants  are used to derive quantum-corrected
dynamical FRW equations. In order to find
the explicit form of the running coupling constants, their (asymptotically
free) behaviour  at high energy scale is used.

The model possesses a de Sitter solution at high curvature to describe
expanding inflationary universe. The bound of de Sitter solution depends on the
value of the running constant of $R^2$-term today. We have demonstrated that
   de Sitter solution is always unstable and takes place near to the Planck
scale. Thus, it is possible to evaluate the instability parameter of the model
and the amplitude of perturbations. The slow-roll conditions are well
satisfied, and the $\eta$ slow-roll parameter is much larger than the
$\epsilon$ slow-roll parameter: their behaviour with respect to the
$e$-folds $N$
seems to be the same of the ones in scalar-tensor theories (see review
~\cite{Faraoni}) for inflation
($\epsilon\sim 1/N^2$ and $|\eta|\sim 1/N$).
The amount of
inflation of the model is sufficiently large, the tensor-to-scalar ratio $r$ is
very close to zero. However, in order to have the correct spectral index
$n_s$
compatible with the Planck data it is necessary to take into account the
contribution of RG improved Gauss-Bonnet term in the action. Note that
other RG-improved surface term ($\Box R$) does not play any
important role during inflation.
At low energy, the effective running constants become constant and we recover the
Friedmann universe.

It would be very interesting to compare the inflationary predictions
(including the exit and reheating)  of
higher-derivative quantum gravity with those of Einstein quantum gravity
in more detail. This will be considered elsewhere.

\section*{Acknoweledgments}
The research by SDO has been supported in part by MINECO(SPAIN), projects
FIS2010-15640 and FIS2013-44881 and by the Russ. Government Program of 
Competitive Growth of Kazan Federal
University.


\begin{thebibliography}{}



\bibitem{WMAP}
G.~Hinshaw {\it et al.}  [WMAP Collaboration],
     Astrophys.\ J.\ Suppl.\  {\bf 208} (2013) 19
     [arXiv:1212.5226 [astro-ph.CO]]; E.~Komatsu {\it et al.}  [WMAP
Collaboration],
     Astrophys.\ J.\ Suppl.\  {\bf 180} (2009) 330
     [arXiv:0803.0547 [astro-ph]];

\bibitem{Planck} P.~A.~R.~Ade {\it et al.}  [Planck Collaboration],
     Astron.\ Astrophys.\  {\bf 571} (2014) A22
     [arXiv:1303.5082 [astro-ph.CO]].

\bibitem{book} V.~Mukhanov,
     ``Physical foundations of cosmology,''
     Cambridge, UK: Univ. Pr. (2005) 421 p;
D.~S.~Gorbunov and V.~A.~Rubakov,
     ``Introduction to the theory of the early universe: Cosmological
perturbations and inflationary theory,''
     Hackensack, USA: World Scientific (2011) 489 p.

\bibitem{Bi} P.~A.~R.~Ade {\it et al.}  [BICEP2 Collaboration],
     Phys.\ Rev.\ Lett.\  {\bf 112} (2014) 241101
     [arXiv:1403.3985 [astro-ph.CO]].

\bibitem{Rich} R.~P.~Woodard,
     arXiv:1407.4748 [gr-qc]; M.~G.~Romania, N.~C.~Tsamis and R.~P.~Woodard,
     Lect.\ Notes Phys.\  {\bf 863} (2013) 375
     [arXiv:1204.6558 [gr-qc]].

\bibitem{Ser}
    M.~Rinaldi, G.~Cognola, L.~Vanzo and S.~Zerbini,
     arXiv:1410.0631 [gr-qc]; K.~Bamba, G.~Cognola, S.~D.~Odintsov and
S.~Zerbini,
     Phys.\ Rev.\ D {\bf 90} (2014) 023525
     [arXiv:1404.4311 [gr-qc]];
    B.~J.~Broy, F.~G.~Pedro and A.~Westphal,
     arXiv:1411.6010 [hep-th];
K.~Bamba and S.~D.~Odintsov,
   Symmetry {\bf 7} (2015) 220
   [arXiv:1503.00442 [hep-th]].


\bibitem{B}I.~L.~Buchbinder, S.~D.~Odintsov and I.~L.~Shapiro,
     Effective action in quantum gravity,
     Bristol, UK: IOP (1992) 413 p;
Riv.\ Nuovo Cim.\  {\bf 12N10} (1989) 1


\bibitem{Sa}
    A.~Salvio and A.~Strumia,
     JHEP {\bf 1406}, 080 (2014)
     [arXiv:1403.4226 [hep-ph]].


\bibitem{test}
J.~Khoury and A.~Weltman,
   Phys.\ Rev.\ D {\bf 69}, 044026 (2004)
   [astro-ph/0309411].



\bibitem{El}
    E.~Elizalde and S.~D.~Odintsov,
     Phys.\ Lett.\ B {\bf 303} (1993) 240;
Phys.\ Lett.\ B {\bf 321} (1994) 199; Z.\ Phys.\ C {\bf 64} (1994) 699
     [hep-th/9401057]; S. D. Odintsov,
     Fortsch.\ Phys.\  {\bf 39} (1991) 621.
\bibitem{rginfl}
A.~De Simone, M.~P.~Hertzberg and F.~Wilczek,
     Phys.\ Lett.\ B {\bf 678} (2009) 1
     [arXiv:0812.4946 [hep-ph]]; H.~M.~Lee,
     Phys.\ Lett.\ B {\bf 722} (2013) 198
     [arXiv:1301.1787 [hep-ph]]; G.~Barenboim, E.~J.~Chun and H.~M.~Lee,
     Phys.\ Lett.\ B {\bf 730} (2014) 81
     [arXiv:1309.1695 [hep-ph]]; N.~Okada and Q.~Shafi,
     arXiv:1311.0921 [hep-ph];  M.~Herranen, T.~Markkanen, S.~Nurmi and
A.~Rajantie,
     Phys.\ Rev.\ Lett.\  {\bf 113} (2014) 21,  211102
     [arXiv:1407.3141 [hep-ph]]; T.~Inagaki {\it et al.},
     arXiv:1408.1270 [gr-qc];  E.~Elizalde, S.~D.~Odintsov, E.~O.~Pozdeeva and
S.~Y.~Vernov,
     Phys.\ Rev.\ D {\bf 90} (2014) 084001
     [arXiv:1408.1285 [hep-th]]; Y.~Hamada, H.~Kawai and K.~y.~Oda,
     JHEP {\bf 1407} (2014) 026
     [arXiv:1404.6141 [hep-ph]; H.~J.~He and Z.~Z.~Xianyu,
     JCAP {\bf 1410} (2014) 019
     [arXiv:1405.7331 [hep-ph]].
\bibitem{Fradkin}
    K.~S.~Stelle,
     Phys.\ Rev.\ D {\bf 16} (1977) 953; E.~S.~Fradkin and A.~A.~Tseytlin,
     Nucl.\ Phys.\ B {\bf 201} (1982) 469;
I.~G.~Avramidi and A.~O.~Barvinsky,
     Phys.\ Lett.\ B {\bf 159} (1985) 269.



\bibitem{L1}
A. Vilenkin, Phys. Rev. D 32 (1985) 2511.

\bibitem{L2}
S. Capozziello, Int. J. Mod. Phys. D114483 (2002).

\bibitem{Monica}
G.~Cognola, M.~Gastaldi and S.~Zerbini,
     Int.\ J.\ Theor.\ Phys.\  {\bf 47}, 898 (2008)
     [gr-qc/0701138].

\bibitem{miolag}
G.~Cognola, L.~Sebastiani and S.~Zerbini,
     arXiv:1006.1586 [gr-qc].
\bibitem{Ei}
M.~B.~Einhorn and D.~R.~T.~Jones,
     arXiv:1410.8513 [hep-th].


\bibitem{Barrow}
J.~D.~Barrow and D.~J.~Shaw,
     Gen.\ Rel.\ Grav.\  {\bf 43}, 2555 (2011)
     [Int.\ J.\ Mod.\ Phys.\ D {\bf 20}, 2875 (2011)]
     [arXiv:1105.3105 [gr-qc]].


\bibitem{corea}
Hwang, J.-C., and Noh, H., Phys. Lett. B, 506, 13--19, (2001);
  H.~Noh and J.~c.~Hwang,
   Phys.\ Lett.\ B {\bf 515}, 231 (2001)
   [astro-ph/0107069].


\bibitem{Bamba}
K.~Bamba, S.~Nojiri, S.~D.~Odintsov and D.~Saez-Gomez,
     arXiv:1410.3993 [hep-th].

\bibitem{GBdisc}
M.~B.~Einhorn {\it et al.}  [ Collaboration],
   arXiv:1412.5572 [hep-th].

\bibitem{Faraoni} Y. Fujii and K.-i. Maeda, The Scalar-Tensor Theory of
Gravitation (Cambridge University Press,
Cambridge, United Kingdom, 2003);
V. Faraoni, Cosmology in scalar-tensor gravity (Springer, 2004).



\end{thebibliography}
\end{document}